\documentclass[a4paper,fleqn,usenatbib]{mnras}

\usepackage{newtxtext,newtxmath}

\usepackage[T1]{fontenc}
\usepackage{ae,aecompl}

\usepackage{graphicx}    
\usepackage{amsmath}    
\usepackage{amssymb}    
\usepackage[normalem]{ulem}
\useunder{\uline}{\ul}{}
\usepackage{booktabs}
\usepackage{bbm}
\usepackage{multirow}

\newcommand{\rms}{$R_{\rm{MS}}$}
\newcommand{\lx}{$L_{\rm X}$}

\title[Binning-free continuous relationship]{A binning-free method reveals a continuous relationship between galaxies' AGN power and offset from main sequence.}

\author[L. P. Grimmett et al.]{
L. P. Grimmett,$^{1}$\thanks{E-mail: lpgrimmett1@sheffield.ac.uk}
J. R. Mullaney,$^{1}$
E. P. Bernhard,$^{1}$
C. M. Harrison,$^{2}$
D. M. Alexander,$^{3}$
\newauthor
\ F. Stanley, $^{4}$
V. A. Masoura, $^{5}$
K. Walters$^{6}$
\\
$^{1}$Department of Physics and Astronomy, University of Sheffield, Sheffield, S3 7RH, UK\\
$^{2}$School of Mathematics, Statistics and Physics, Newcastle University, Newcastle upon Tyne, NE1 7RU, UK\\
$^{3}$Centre for Extragalactic Astronomy, Department of Physics, Durham University, South Road, Durham, DH1 3LE, UK\\
$^{4}$Department of Space, Earth and Environment, Chalmers University of Technology, Onsala Space Observatory, SE-43992 Onsala, Sweden\\
$^{5}$National Observatory of Athens, V. Paulou \& I. Metaxa, 15236 Penteli, Greece\\
$^{6}$School of Mathematics and Statistics, University of Sheffield, Sheffield, S3 7RH, UK
}

\date{Accepted XXX. Received YYY; in original form ZZZ}

\pubyear{2017}

\begin{document}
\label{firstpage}
\pagerange{\pageref{firstpage}--\pageref{lastpage}}
\maketitle

\begin{abstract}
Studies investigating the relationship between AGN power and the star formation rates (SFRs) of their host galaxies often rely on averaging techniques -- such as stacking -- to incorporate information from non-detections. However, averages, and especially means, can be strongly affected by outliers and can therefore give a misleading indication of the ``typical'' case. Recently, a number of studies have taken a step further by binning their sample in terms of AGN power (approximated by the 2-10keV luminosity of the AGN), and investigating how the SFR distribution differs between these bins. These bin thresholds are often weakly motivated, and binning implicitly assumes that sources within the same bin have similar (or even identical) properties. In this paper, we investigate whether the distribution of host SFRs -- relative to the locus of the star-forming main sequence (i.e., \rms) -- changes \textit{continuously} as a function of AGN power. We achieve this by using a hierarchical Bayesian model that completely removes the need to bin in AGN power. In doing so, we find strong evidence that the \rms\ distribution changes with 2-10keV X-ray luminosity. The results suggest that higher X-ray luminosity AGNs have a tighter physical connection to the star-forming process than lower X-ray luminosity AGNs, at least within the $0.8<z<1.2$ redshift range considered here.
\end{abstract}




\section{Introduction}

The proportion of galaxies that show evidence of Active Galactic Nuclei (i.e., AGN) ranges from a few percent to a few tens of percent, depending on galaxy mass \cite[e.g.,][]{Kauffmann03, Best05, Mullaney12jan, Kaviraj19}. What this implies is that an individual supermassive black hole (SMBH) spends most of cosmic time in a “dormant” state during which it accretes at such a low rate as to make it unidentifiable as an AGN \cite[e.g.][]{Heckman04}. What is clear from their high masses, however, is that all SMBHs – irrespective of their current accretion rate – must have undergone periods of rapid growth at earlier times \cite[e.g.][]{Soltan82}. Since BH growth is not a constant, it raises the question of what external factors cause a SMBH to transition from a dormant state to an active state (and vice versa). Or, more succinctly, what galaxy properties, if any, dictate AGN power?

Recent observations of the inner few (tens of) parsecs of galaxies hosting AGNs have revealed evidence of bars and spiral structures that may be funnelling material toward the central SMBH \cite[e.g.][]{Shlosman89, Storchi-Bergmann07, Audibert19, Shimizu19}. While such studies are important for revealing how gas and dust are transferred from the host galaxy, they do not address the question of what ``macroscopic'' galaxy properties help to trigger black hole growth. This is important because, since the energy released by AGNs is thought to impact on galaxy scales, it is crucial that we understand what large-scale galaxy properties make them susceptible to triggering SMBH growth.

A key means of investigating what galaxy-scale factors govern SMBH growth rates is by quantifying the properties of AGN-hosting galaxies and attempting to identify correlations between these host properties and AGN power. However, this is hampered by the fact that, compared to most other galactic processes (e.g., star-forming events, mergers), AGNs are extremely variable and short-lived. As demonstrated by \cite{Hickox14}, this stochastic duty cycle tends to dilute the underlying connections between AGN power and other galactic properties, such that plots of mean galaxy star formation rate (SFR) vs. AGN power, for example, show a flat (i.e., independent) relationship \cite[e.g.][]{Harrison12, Rosario12, Mullaney12jan, Stanley15, Stanley17, Suh17, Ramasawmy19}. Recently, some studies have instead investigated the {\it distribution} of star forming properties (as opposed to simple means) in bins of AGN power \cite[e.g.][]{Scholtz18,Bernhard19}. Specifically, \cite{Scholtz18} compared the distribution of specific SFR in two X-ray luminosity (\lx ) bins, but did not find any significant evidence of a difference between the two bins ($43 < \log_{10}(L_{\rm X}/ \text{ergs s}^{-1}) < 44$ and $44 < \log_{10}(L_{\rm X}/\text{ergs s}^{-1}) < 45$). \citet[][B19 from hereon in]{Bernhard19} compared the distribution of the \rms\ statistic (\rms\ $=\rm{SFR}/\rm{SFR}_{\rm MS}$, where $\rm{SFR}_{\rm MS}$ {is the expected SFR for a galaxy of identical mass and redshift, should it lie on the locus of the star-forming main sequence) in bins of low \lx\ (i.e., $42.53 < \log_{10}(L_{\rm X}/\text{erg s}^{-1}) < 43.3$) and high \lx\ (i.e., $43.3 < \log_{10}(L_{\rm X}/\text{erg s}^{-1}) < 45.09$), and only found ``tentative'' evidence of a dependency. Note that, the \rms\ statistic is often referred to as the ``starburstiness'' of a galaxy \citep{Elbaz11, Schreiber15}; we shall also use this notation throughout the rest of the paper.

So whilst the use of distributions has allowed us to investigate the star-forming properties of AGNs in more detail than using simple averages, no study has demonstrated that the distribution of star-forming properties is dependent on $L_{\rm X}$.\footnote{Note, here we use ``dependence'' in the strict mathematical sense, rather than suggesting that SFR physically depends on AGN power.} Of course, this may be because no intrinsic connection exists. It could, however, be due to an often unaddressed limitation in the analysis: the use of arbitrarily-constructed bins of \lx . Beyond being somewhat arbitrary, weakly-motivated and possibly impacting results \citep{Lanzuisi17}, binning has several further limitations. One problem is how to classify a source which, when considering errors, could fall in two or more bins (i.e., if there was a bin boundary at $\log_{10}L_{\rm X}=44$, can a source with $\log_{10}L_{\rm X}= 43.8 \pm 0.3$ be accurately classified?). In an attempt to deal with sources that, within their errors, cross a bin boundary, some studies choose to discard ambiguous sources \cite[e.g.][]{Grimmett19}, whereas other studies assign them to a particular bin based only on their measured value \cite[e.g.][]{Delvecchio15, Aird18a, Bernhard19}. Both of these approaches have intrinsic problems. Information is lost when sources are just discarded and by just using the measured value, uncertainties in the binning axis are not fully appreciated. A second limitation is the implied assumption that all sources in the same bin have identical properties, yet sources just either side of the bin boundaries are different. Both of these limitations constitute a loss of information from the data in hand.

In this study, to investigate the implications of binning on our investigations of the relationship between star-forming properties and AGN power, we analyse the \rms \ distribution as a continuous function of $L_{\rm X}$. To do this, we have developed a comprehensive Bayesian hierarchical model which has two substantial benefits over binning. Firstly, it allows us to eliminate the possibility of binning-dependent results. Secondly, the model allows us to accurately account for \text{all} uncertainties (including, where necessary, upper limits) on the independent variable (i.e., in our case \lx ). \footnote{Multiple dependent data sources can easily be adopted in to the framework, but for this study we choose only to model $L_{\rm X}$ as a demonstration of the technique.} Specifically, this paper aims to quantify the dependence between the \rms \ distribution and $L_{\rm X}$, without the need for binning or averaging. In doing so, we extract all available information from our data and find strong evidence of a relationship between the star-forming properties of AGN-hosting galaxies and \lx .

The outline of the paper is as follows. In Section~\ref{data} we briefly summarise how the dataset was constructed. In Section~\ref{model} we summarise the hierarchical Bayesian model, explain how we eliminate the need for binning and briefly introduce our MCMC model switching algorithm, which will test whether the \rms\ distribution is dependent on \lx. In Section~\ref{results} we present the output of the analysis and discuss the limitations and implications in Section~\ref{discussion}. Where necessary, we adopt a WMAP-7 year cosmology \cite{Larson11} and assume a \cite{Chabrier03} initial mass function. Finally, in Appendix~\ref{app:mcmc} we give the full details of the MCMC model switching algorithm.

\section{Data} \label{data}

So that we can compare the results of our new method with previously found results, we decide to reuse the same dataset as constructed in B19. This will ensure that any differences are the direct result of the analysis method, rather than from differences between two independent data sets. We provide a summary of the sample derivation in this section, but refer interested readers to B19 for a fuller explanation.

Briefly, we take the 541 X-ray detected sources with a redshift between $0.8 < z \leq 1.2$ from the \textit{COSMOS Legacy Survey} \citep{Civano16, Marchesi16}. This small redshift range ($\sim75$ per cent have spectroscopic redshifts) is chosen to minimise any potential redshift effects. These sources have rest-frame 2-10 keV, absorption-corrected X-ray luminosities spanning the range $42.53 < \log_{10}(L_{\rm X}/\text{erg}~s^{-1}) < 45.09$ (see \citealt{Marchesi16} for details on how they calculated \lx , including how they corrected for absorption). We should note that in order to remain consistent with B19 for the aforementioned purposes, we do not include those sources with upper limits on $L_{\rm X}$ nor account for redshift variation,  although it would be straightforward to do so as explained in Section~\ref{priorlx}. Uncertainties on \lx\ values are derived by converting the percentage error on the flux measurement presented in \cite{Marchesi16}. On comparing these errors to the upper and lower \lx\ bounds in \cite{Marchesi16}, we find that our uncertainties are generally more conservative. We then derive a SFR for each source using the DECOMPIR code (see \citealt{Mullaney11} for full details) on the super-deblended photometry presented in the catalogue of \cite{Jin18}which used the deblending technique of \cite{Liu18}. The catalogue contains data from various sources such as \textit{Spitzer} and \textit{Herschel} and covers the 24-1200$\mu$m range.

In total, our sample contains 148 AGNs with measured SFRs, and 393 with upper limits on their SFRs. Stellar masses are calculated using the multi-wavelength spectral energy distribution fitting code CIGALE \citep{Noll09, Serra11, Ciesla15, Boquien19}, using the same parameter prescription as \cite{Grimmett19}. The stellar mass parameters were chosen to maximise the accuracy according to the testing presented in \cite{Ciesla15}. Next, we use the prescription of \cite{Schreiber15}, together with each galaxy's redshift and mass, to predict the SFR that it would have if it were on the star-forming main sequence (i.e., $\text{SFR}_{\rm{MS}}$). Finally, we calculate the ``starburstiness'' statistic, \rms , of each galaxy in our sample via $R_{\rm MS}=\frac{\rm SFR}{\rm SFR_{MS}}$. The \rms\ value of a galaxy aims to provide an indication of the star-forming properties of a galaxy after taking into account the mass and redshift dependence of the SFR of the dominant population of so-called main sequence galaxies \cite[e.g.][]{Brinchmann04,Noeske07,Elbaz07,Magdis10, Schreiber15}. While we appreciate that the precise nature of the mass and redshift of the main sequence is still the matter of some debate \cite[e.g.][]{Speagle14, Ilbert15, Whitaker15,Popesso19a}, the main aim of this study is to demonstrate a new analysis technique, so we choose to the use the definition of \cite{Schreiber15} to remain consistent with B19. This again ensures that any differences in results are a direct consequence of the analysis technique, as oppose to differences in the sample.

\section{The continuous model, model selection and MCMC algorithm} \label{model}

In this section we describe how we model the \rms \ data, in such a way to remove the need for binning, which enables us to investigate whether (and, if so, how) the \rms \ distribution changes as a continuous function of $L_{\rm X}$. In subsection \ref{rms_distrib}, we introduce the log-normal distribution we use to model the \rms\ distribution and explain why we must use a ``hierarchical'' Bayesian approach to allow this to vary continuously with \lx . Next, in subsection \ref{priorandpost} we describe our Bayesian priors and how these provide a mechanism to include all uncertainties on each individual \lx \ value. Finally, in subsection \ref{mcmc}, we introduce our bespoke MCMC sampler that explores the posterior parameter space in a way that allows us to test whether the \rms\ distribution depends on \lx.

\subsection{$R_{\rm{MS}}$ distribution and likelihood function}
\label{rms_distrib}

In order to test the continuous relationship between the \rms \ distribution and \lx\ we assume a functional parametric form for the \rms \ distribution. In this work, we choose to model the \rms \ distribution as a log-normal distribution (i.e., that $\log_{10} (R_{\rm{MS}})$ is normally distributed). A log-normal distribution is primarily chosen to remain consistent with B19. Although recent studies have found the scatter around the main sequence to be well modelled by a log-normal distribution \citep{Rodighiero11, Sargent12, Guo13, Chang15, Mullaney15, Caplar19, Davies19} there is likely a ``bump'' in the high-\rms\ end of the distribution caused by starburst galaxies. Indeed, it is also true that there is likely an additional component at lower \rms\ values due to the population of quiescent galaxies. Therefore the accuracy of using a log-normal distribution could be questioned. However, we leave devising a more flexible model for a future work, where we intend to include all three populations (i.e., quiescent, main sequence and starburst galaxies) in our model (Grimmett et al., in prep). Therefore we stress that this study is working under the assumption that the deviation from the main sequence of star formation is log-normally distributed, at least for AGNs. In future studies, this model could be made more flexible to account for an additional second component, but the primary motivation of this work is to test the ability of the method to remove the need for binning and therefore we choose a log-normal \rms\ distribution to remain consistent with B19.

As we choose to use a Bayesian approach, we wish to derive the posterior distribution, which is proportional to the product of the data-driven likelihood function (assuming a log-normal \rms\ distribution) and the prior distributions. We are then interested in sampling parameter values from this posterior distribution. The prior distributions are essential for including the uncertainty on \lx \ and are fully explained in Section~\ref{priorandpost}. The remainder of this section, therefore, describes how we derive the likelihood function.

The likelihood function is given by the product of the probability density functions (PDFs) of all the detected \rms \ values, and the cumulative distribution functions (CDFs) of all undetected sources. The PDF of a given detected $R_{\rm{MS},i}$ value with parameters $\mu$ (representing the mode) and $\sigma$ (representing the width), is given by

\begin{equation}
    \label{eq:pdf}
    f(\log_{10}(R_{\rm{MS}, i})|\mu, \sigma) = (2\pi\sigma^2)^{-\frac{1}{2}} \exp \left( -\frac{(\log_{10}(R_{\rm{MS}, i}) - \mu)^2}{2\sigma^2}\right).
\end{equation}
For upper limits (i.e., non-detected \rms \ values, which ultimately comes from an upper limit on the infrared flux) the PDF is replaced by the CDF. The CDF is the integral of the PDF and can therefore be written as,

\begin{equation}
    \begin{split}
    F(\log_{10}(R_{\rm{MS}})|\mu, \sigma) &= \int_{-\infty}^{R_{\rm{MS}}} f(X|\mu, \sigma) dX\\ &= \frac{1}{2}\left(1 + \text{erf}\left( \frac{\log_{10}(R_{\rm{MS}}) - \mu}{\sigma\sqrt{2}}\right)\right),
    \end{split}
\end{equation}
where $f(X|\mu, \sigma)$ is given by Equation~\ref{eq:pdf}.

In other words, for a given galaxy, $F(\log_{10}(R_{\rm MS}))$ is close to 1 if most of the \rms\ distribution with given $\mu$ and $\sigma$ values lies below the value of the upper limit. By contrast, $F(\log_{10}(R_{\rm MS}))$ is close to 0 if most of the distribution lies above the upper limit, meaning those $\mu$ and $\sigma$ values are {\it incompatible} with that limit.

By combining both our $m$ detections, $R_{\rm{MS, 1}}, ..., R_{\rm{MS, m}}$, and $n-m$ non-detections, $R_{\rm{MS, m+1}}, ..., R_{\rm{MS, n}}$, the likelihood function is given by the product of the PDFs (for the detections) and the CDFs (for the upper limits),

\begin{equation}
    \begin{split}
    L(\log_{10}(R_{\rm{MS}})| \mu, \sigma) = &\prod_{i=1}^m f(\log_{10}(R_{\rm{MS, i}})|\mu, \sigma)\\ &\prod_{i=m+1}^n F(\log_{10}(R_{\rm{MS, i}}))|\mu, \sigma).
    \end{split}
\end{equation}

If we were going to assume no dependence of \rms \ on \lx , and no uncertainty on \lx , then at this stage we could simply find the best-fitting values for $\mu$ and $\sigma$, as has been used previously in ``Bayesian''-style studies that use bins. Such studies derive the likelihood function in different bins, use parameter-maximisation techniques to find the best fitting value for $\mu$ and $\sigma$ within each bin, and then compare how parameters change between different bins \citep[e.g.][]{Mullaney15, Scholtz18, Bernhard19}. However, in order to analyse the \rms \ distribution as a continuous function of $L_{\rm X}$, we {\it must} use a hierarchical model, since this allows the parameters that control the shape of the \rms\ distribution (i.e., $\mu$, $\sigma$) to vary as a function of $L_{\rm X}$. As the true relationship between the $\mu$ and $\sigma$ parameters and the \lx\ values is unknown, the choice of relationship is arbitrarily specified. However, in order to test the case of no dependence (i.e., that \rms \ and \lx\ are independent of one another), it is sufficient to show that a simple model that allows dependence is preferable to one that imposes independence. Therefore, we choose to use simple functions to relate the parameters of the \rms \ distribution and the \lx\ values (hereafter referred to as the ``functional relationships''), given by:

\begin{equation}\label{muform}
    \mu_i = \theta_0 + \theta_1 \log_{10}\left(\frac{L_{\rm {X},i}}{10^{40}}\right) \ \ \ \text{and} \ \ \
    \sigma_i = e^{\theta_2 + \theta_3 \log_{10}\left(\frac{L_{\rm {X},i}}{10^{40}}\right)}.
\end{equation}
The rescaling of the \lx \ values ensures that the hyperparameters are not orders of magnitude different, which could lead to problems in the analysis. Note that, throughout this paper, we are only considering the effect of \lx\ on the \rms\ distribution and hence our functional relationships only factor-in \lx. If other parameters, such as redshift or stellar mass were also to be considered, they could be added to the functional relationships as described in Equation}~\ref{muform}. Such an expansion of the model is, however, beyond the scope of the current study, but would remove the need for binning in both redshift and stellar mass.

By introducing these functional relationships, we have essentially related the mode and width of the \rms \ distribution to the \lx \ values. Additionally, we have changed the parameters of interest from $\mu$ and $\sigma$ to the parameters given by $\boldsymbol{\theta} = \{\theta_0, \theta_1, \theta_2, \theta_3\}$ (hereafter, our hyperparameters); this is what makes the approach ``hierarchical''. Note that we specify an exponential form for the functional relationship between $\sigma_i$ and $L_{\rm {X},i}$ as $\sigma_i$ cannot be negative. The focus of this analysis is to now find the posterior distributions for $\boldsymbol{\theta}$. By considering these posteriors, the functional relationships allow us to test whether the \rms\ distribution is dependent upon \lx. For example if $\theta_1 = \theta_3 = 0$, the functional relationships are no longer a function of \lx \ and therefore imply that the \rms \ distributions are independent of $L_{\rm X}$. Additionally, relating the mode and width of the \rms \ distribution to the \lx values has completely removed the need to bin the data in \lx. The question of independence now becomes how likely is $\theta_1 = \theta_3 = 0$, given the data observed. More details of which are contained in Section~\ref{switch}.

As a result of adapting the mode and width of the distribution so that binning is not required, the likelihood function changes slightly and is now given by,
\begin{equation}
    \begin{split}
    L(\boldsymbol{\theta}, L_{\rm X}| R_{\rm{MS}}) = & \prod_{i=1}^m f(\log_{10}(R_{\rm{MS, i}})|\boldsymbol{\theta}, L_{\rm{X,i}})\\ & \prod_{i=m+1}^n F(\log_{10}(R_{\rm{MS, i}})|\boldsymbol{\theta}, L_{\rm{X,i}}).
    \end{split}
    \label{lik_function}
\end{equation}

\subsection{Prior and posterior distributions}\label{priorandpost}

\subsubsection{Prior distribution on $L_{\rm X}$} \label{priorlx}

We have now expressed the parameters as functions of the independent data (in this case, $L_{\rm X}$) and the hyperparameters, $\boldsymbol{\theta}$. The next step we must now consider is how to fully account for uncertainties on $L_{\rm X}$. In our hierarchical model, we are able to treat the \lx\ values as parameters, and can therefore place informative Bayesian priors on their values. The prior distribution on each $L_{\rm X,i}$ can be constrained by the measured value $L_{\rm {X,i, meas}}$ and uncertainty $\xi_i$ and modelled as a log-normal (here, we are assuming that our errors are symmetric in log space).  This means that the prior distribution on a specific $\log_{10}(L_{\rm {X,i}})$ is given by,

\begin{equation}
    \begin{split}
    f(\log_{10}(L_{\rm {X,i}})|&\log_{10}(L_{\rm {X,i, meas}) }, \xi_i) =\\
    &(2\pi\xi_i)^{-\frac{1}{2}}\exp \left(-\frac{(\log_{10}(L_{\rm {X,i}}) - \log_{10}(L_{\rm {X,i, meas} }) )^2}{2\xi_i^2} \right).
    \end{split}
\end{equation}
where $\xi_i$ is derived by converting the percentage error on the flux measurement presented in \cite{Marchesi16}. This can be thought of as the probability density of observing the true $L_{\rm X}$ given we have observed a measurement, $L_{\rm {X,i, meas}}$ and error $\xi_i$. It should be noted that in this study we are working with only detected X-ray luminosities to remain consistent with B19 and we assume all uncertainties are modelled with a log-normal. One could, however, replace this prior distribution with any probability distribution. Note that in this study, we have not accounted for the uncertainties on the \rms\ values. This is largely to remain consistent with the modelling approach of B19. In future studies, uncertainties on the dependent variable (in our case, \rms) can be included using a similar method as the one applied to the uncertainties on \lx. Whilst we do not believe that excluding these uncertainties has a major impact on our results, it is a limitation of this study. However, it is not a limitation of the methodology.

At this stage, we have specified our likelihood function (Equation~\ref{lik_function}) and our priors on \lx. The final terms we must consider are the prior distributions on the hyperparameters, which we discuss in the next subsection.

\subsubsection{Prior distribution on hyperparameters} \label{switch}

Because our primary scientific aim is to determine {\it whether} the \rms\ distribution changes with \lx , we are most interested in the (posterior) probability that the hyperparameters $\theta_1$ and $\theta_3$ are equal to 0 or whether they are non-zero (i.e., there is a dependence on \lx).  We therefore choose the prior distributions of these hyperparameters to be a ``spike and slab distribution''. This type of prior allows us to join two distributions; one defined in discrete space (the spike) and one in continuous space (the slab). This is necessary so that we can ensure that there is a defined prior probability that $\theta_1 = 0$ and $\theta_3 = 0$ (i.e., there is a prior probability of independence between \rms \ and \lx), as oppose to a just a probability density. If we have a defined prior probability then we can calculate a posterior probability, again as opposed to just to a probability density. \footnote{A probability density is a ``relative'' likelihood as opposed to an absolute one. For a distribution over a continuous space, the absolute probability of any one particular occurrence is 0, whilst the probability density can be non-zero. For a distribution over a discrete space, the probability mass function (the discrete equivalent of the density) is an absolute probability.}

Our spike and slab prior distributions take the form,

\begin{equation}
    \begin{split}
    &f(\theta_1|\omega) = (1-\omega) \text{N}(\theta_1; {\rm mean}=0, {\rm S.D.}=1) + \omega\delta_{\theta_1=0},\\
    &f(\theta_3|\omega) = (1-\omega) \text{N}(\theta_3; {\rm mean}=0, {\rm S.D.}=1) + \omega\delta_{\theta_3=0},
    \end{split}
\end{equation}
where  $\omega$ is the prior probability that $\theta_1, \theta_3 = 0$ and  $\delta_{\theta_i=0}$ is the delta function. For our analysis, we choose $\omega = 0.5$ so that our prior probability favours neither the case of independence, $p(\theta_1 = 0) = p(\theta_3 = 0) = 0.5$, nor  the case of dependence $p(\theta_1 \neq 0 = p(\theta_3 \neq 0) = 0.5$. As we are not interested in the posterior probabilities that $\theta_0, \theta_2 = 0$, the prior distributions on these parameters are Gaussian distributions with mean 0 and standard deviation 1.

This means that by using spike and slab prior distributions we have constructed four potential models:

\begin{itemize}
    \item{Model 1: $\theta_1 = 0, \theta_3 = 0$, no dependence on $L_{\rm X}$ at all}
    \item{Model 2: $\theta_1 \neq 0, \theta_3 = 0$, mode depends on $L_{\rm X}$, width does not}
    \item{Model 3: $\theta_1 = 0, \theta_3 \neq 0$, width depends on $L_{\rm X}$, mode does not}
    \item{Model 4: $\theta_1 \neq 0, \theta_3 \neq 0$, both mode and width depend on $L_{\rm X}$.}
\end{itemize}

Note that as we have chosen $\omega = 0.5$ our prior distributions give no preferential weight to any of the model scenarios (according to the prior, they all have a probability of 0.25). Having now derived the likelihood function and all needed prior distributions we can construct the final posterior distribution,

\begin{equation}
    \begin{split}
    f(\boldsymbol{\theta}, \mathbf{\log_{10}(L_{\rm X})}|&\mathbf{\log_{10}(R_{MS}}),\mathbf{\log_{10}(L_{\rm {X, meas}})})=\\
    &L(\mathbf{\log_{10}(R_{MS})} | \boldsymbol{\theta}, \mathbf{\log_{10}(L_{\rm X})}) \\
    & \times f(\mathbf{\log_{10}(L_{\rm {X}})}|\mathbf{\log_{10}(L_{\rm {X, meas} })}, \boldsymbol{\xi}) \\
    & \times f(\boldsymbol{\theta}|\omega)\\
    \end{split}
\end{equation}

\subsection{MCMC algorithm and model switching} \label{mcmc}

As our posterior distributions cannot be derived analytically, we have written a purpose-built MCMC sampler in order to sample from the posterior distributions of each given hyperparameter (i.e., $\theta_0$, $\theta_1$, $\theta_2$, $\theta_3$). However, in addition to sampling from the posterior distributions to find the most likely hyperparameter values, we also use our sampler to determine the posterior probability of each of our four models (i.e., for model comparison). The posterior probability of the models can be calculated analytically, however even advanced sampling methods (e.g. Nested Sampling, see \citealt{Buchner14}) struggle to accurately calculate them due to the high dimensionality of our parameter space (i.e., up to 545 dimensions as a result of including the \lx\ values as parameters). Instead, we use ``model switching'' to compute the posterior model probabilities. In this subsection, we summarise our MCMC sampler, including the model switching component; a full description is, however, given in Appendix~\ref{app:mcmc}.

For the most part, our MCMC sampler adopts a standard Metropolis-Hastings (MH) algorithm \citep{Metropolis53, Hastings70} to explore the parameter space. On each iteration, the MH algorithm proposes a new set of parameter values, which are then accepted or rejected. For efficiency, we propose new values for two parameters at a time, and accept them based on their ``acceptance ratio'' (see Equation~\ref{ar}). Our parameter vector is given by $\boldsymbol{\theta} = (\theta_0, \theta_1, \theta_2, \theta_3, \log_{10}(L_{\rm{X}, 1}), ..., \log_{10}(L_{\rm{X}, 1}))$ and therefore we sample $\theta_0, \theta_1$ together and $\theta_2, \theta_3$ together. This is important as the value of $\theta_0$ is dependent on the value of $\theta_1$; similarly, the value of $\theta_2$ is dependent on $\theta_3$. Proposing the dependent hyperparameters together can allow us to take into account the dependency and therefore propose more sensible values.

If we were only considering one model, and simply wished to sample the posterior distributions, then we would simply iterate the above process. However, in our case we wish to compare the relative probability of four different models. As mentioned above, we do this using a technique known as ``model switching'', which we describe next. For the purposes of this explanation, we will assume that the current state of the MCMC algorithm is such that it is in Model 1 (i.e., $\theta_1 = \theta_3 = 0$; however, for simplicity we will ignore $\theta_3$ for the rest of this explanation). We then propose, with probability 0.5, that the new value of $\theta_1$ remains at zero. If it does, we remain within Model 1 and the MH algorithm progresses as usual.

If, however, the new value of $\theta_1$ is chosen to be non-zero, then this implies that the MCMC algorithm has proposed a switch to a different model (in this case, Model 2). If this happens, we cannot retain the value for $\theta_0$, since the value of $\theta_0$ in Model 1 is likely very different to the value of $\theta_0$ in Model 2, as we are now including the $\theta_1$ parameter. This means that, when we propose a model switch, we cannot simply keep $\theta_0$ as before, as it is unlikely to be in a region of high posterior probability. Therefore, we need to propose ``reasonable'' values for both $\theta_0$ and $\theta_1$, given that we have proposed Model 2.\footnote{How we obtain a ``reasonable'' values is explained in full in Appendix~\ref{app:mcmc}} Given the two new proposed values (i.e., $\theta' = (\theta_0',\theta_1')$), the acceptance probability, $\alpha$ is given by,

\begin{equation}
    \label{ar}
    \alpha = {\rm min}\left(\frac{\pi(\theta')q(\theta', \theta)}{\pi(\theta)q(\theta,\theta')}, 1 \right),
\end{equation}
where $\pi(\theta)$ is the full conditional of $\theta$ and $q(\theta,\theta')$ is the proposal density (i.e., the probability density of proposing $\theta'$ given the current $\theta$). Usually, the proposal density is a symmetric function (e.g. a Gaussian), so $q(\theta,\theta')=q(\theta',\theta)$ and the two $q$ values cancel in Equation. \ref{ar}. However, as we explain in Appendix \ref{app:mcmc}, this is not the case when we propose a switch between models \citep{Gottardo08}. We also explain in Appendix \ref{app:mcmc} how we calculate the values for $q(\theta,\theta')$ and $q(\theta',\theta)$. The final stage is the same whether we have proposed a model switch or not: we accept the proposed values with probability equal to the acceptance ratio, otherwise we re-accept the current values (as is standard in an MH algorithm).

The above process is replicated for $\theta_2$ and $\theta_3$ (in this case, a change from $\theta_3=0$ to $\theta_3\neq0$, or vice versa, represents a switch between models) and then the sampler works through the rest of the parameter vector, individually. The process is more straightforward for the $L_{\rm X}$ values as the proposal distribution is centered on the current value and no switching is required, so our process reverts to the standard MH sampler. As we describe fully in our Appendix, by the construction of our MCMC algorithm, the models that we can switch to depends on the current model. For example, if the chain is currently in Model 1 it cannot jump to Model 4, as that would require a change in the dependency on $\mu$ and $\sigma$ at the same time, whereas we only consider changes in these dependencies one at a time. Again, by construction, we have ensured that over the entire chain, all models are proposed equally (i.e, with a probability of 0.25).

In one iteration we sample through the full parameter vector and we run five chains in parallel for 25,000 iterations.\footnote{The choice of five chains for 25,000 iterations is arbitrary, but these values ensured that the combined chain contained a sufficiently high number of samples from the posterior.} Each chain has the first 5000 iterations removed as a burn-in, then the remaining iterations from each chain are combined to form the final sample of 100,000 posterior draws for each parameter. The posterior probability of each of the four models presented in Section~\ref{switch} is then straightforward to calculate from the combined chain: all we need to do is calculate the fraction of accepted samples from each model in the combined chain.

\section{Results} \label{results}

Given that we now have 100,000 independent draws from the posterior distribution from each parameter, we can begin to investigate the relationship between the \rms \ distribution and \lx. Recall that we modelled the \rms \ distribution as a log-normal distribution and set a relationship between the mode and width, and the \lx \ values as outlined in Equation~\ref{muform}. We proposed values such that our sample was forced to consider $\theta_1 = 0$ and $\theta_3=0$ respectively, effectively allowing for the MCMC sampler to switch between models of dependence or independence. In this Section, we present the posterior distributions of the hyperparameters and the posterior model probabilities.

\subsection{Posterior distributions}

\subsubsection{Posterior model probabilities} \label{sec:postmodel}

\begin{table*}
    \centering
    \begin{tabular}{|c|c|c|c|c|}
    \hline
    Model & Value of $\mu$                                                        & Value of $\sigma$                                                          & Posterior probability & Bayes Factor vs. Model 1 \\ \hline
    1     & $\theta_0$                                                            & $e^{\theta_2}$                                                          & 0.06102               & -                   \\
    2     & $\theta_0 + \theta_1 \log_{10}\left(\frac{L_{\rm X}}{10^{40}}\right)$ & $e^{\theta_2}$                                                          & 0.00477               & 0.0781                \\
    3     & $\theta_0$                                                            & $e^{\theta_2+\theta_3 \log_{10}\left(\frac{L_{\rm X}}{10^{40}}\right)}$ & 0.00148                & 0.02425               \\
    4     & $\theta_0 + \theta_1 \log_{10}\left(\frac{L_{\rm X}}{10^{40}}\right)$ & $e^{\theta_2+\theta_3 \log_{10}\left(\frac{L_{\rm X}}{10^{40}}\right)}$ & 0.93273               & 15.285               \\
    \end{tabular}
    \caption{The posterior model probabilities given for each model. These are calculated by considering the amount of time the MCMC chain spent in each of the models. Also shown is the Bayes Factor, which is used to judge, out of two models, the model considered to be the most likely.}
    \label{tab:model1}
\end{table*}

As a result of implementing model switching in the MCMC algorithm we can easily calculate the posterior model probabilities by considering the fraction of samples of each chain within each model. The posterior model probabilities alongside the Bayes Factor comparison to the independent Model 1 are given in Table~\ref{tab:model1}. The Bayes Factor, which can be accurately used to compare two models \citep{Kass95}, is given as the ratio of the posterior model probability of the more complex model to the posterior model probability more simple one. Naturally, the Bayes Factor includes a ``penalty'' for the number of parameters used. In our case, as a result of including \lx \ values as a parameters our models have vastly different numbers of parameters. Model 1, which ignores \lx \ values only has 2, whereas Models 2, 3 and 4 have 544, 544 and 545 respectively. This can help explain the very small posterior probabilities of Models 2 and 3, where the chain either prefers the simple Model 1, or for the sake of 1 extra parameter Model 4, which comprehensively outperforms them. The Bayes Factor comparing Model 4 to Model 1 gives us a value of 15.285, which can be seen as ``strong'' evidence in favour of Model 4 \citep{Kass95}. Using this model comparison model technique, the posterior model probability is not equal to the probability that the model is true, as the sum of all posterior model probabilities in the analysis must be equal to 1. It is therefore important to consider the Bayes Factor approach for comparing the models, rather than using the posterior model probabilities as they are.

\subsubsection{Hyperparameters}

\begin{figure}
    \includegraphics[width=\columnwidth]{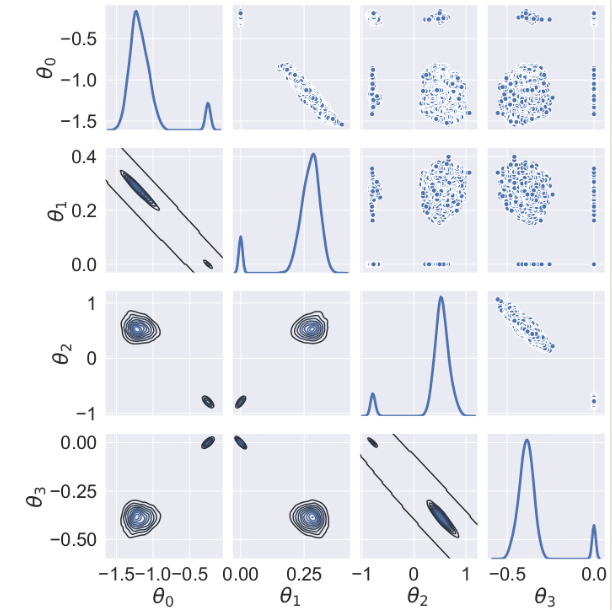}
    \caption{The output from our MCMC algorithm. The on-diagonal plots show the marginalised posterior distributions for each parameter, with the joint posterior distributions shown by the off-diagonal contour plots. The figures include results from the entire MCMC chain, which means that different peaks (on-diagonal) and contour regions (off-diagonal) illustrate when the chain is in a particular model. For example, in the plot in the second row, first column (from top left), the larger of the two contour regions corresponds to $\theta_1 \neq 0$, which is the case in both Model 2 and Model 4. From this posterior plot alone, one cannot distinguish whether the chain is in Model 2 or Model 4, as information about the other parameters is needed (i.e., a 4-dimensional plot would show four discrete model regions). Secondly, there is a smaller region in the lower-right corner that corresponds to the region where $\theta_1=0$, which is the case for both Model 1 and Model 3. Again, one cannot distinguish between these two models from this plot alone. However, given the negligible amount of time the chain spends in Model 2 and Model 3, it can be assumed without much loss of accuracy that the larger region represents the likelihood for Model 4 and the smaller region represents the likelihood for Model 1. This is analogous to the larger and smaller peaks in the on-diagonal plot for $\theta_1$.}
    \label{posterior}
\end{figure}

In Figure~\ref{posterior} we present the posterior distributions for the hyperparameters as computed by the MCMC algorithm outlined in Section~\ref{mcmc}. The off-diagonal plots show the joint posterior distributions. As described in Section~\ref{sec:postmodel}, we have strong evidence that a model of the \rms \ distribution with a dependence on \lx \ is preferred to the independent model. The rest of this paper therefore, works with the assumption that Model 4 is the most suitable model.

\begin{figure}
    \includegraphics[width=\columnwidth, height=15cm]{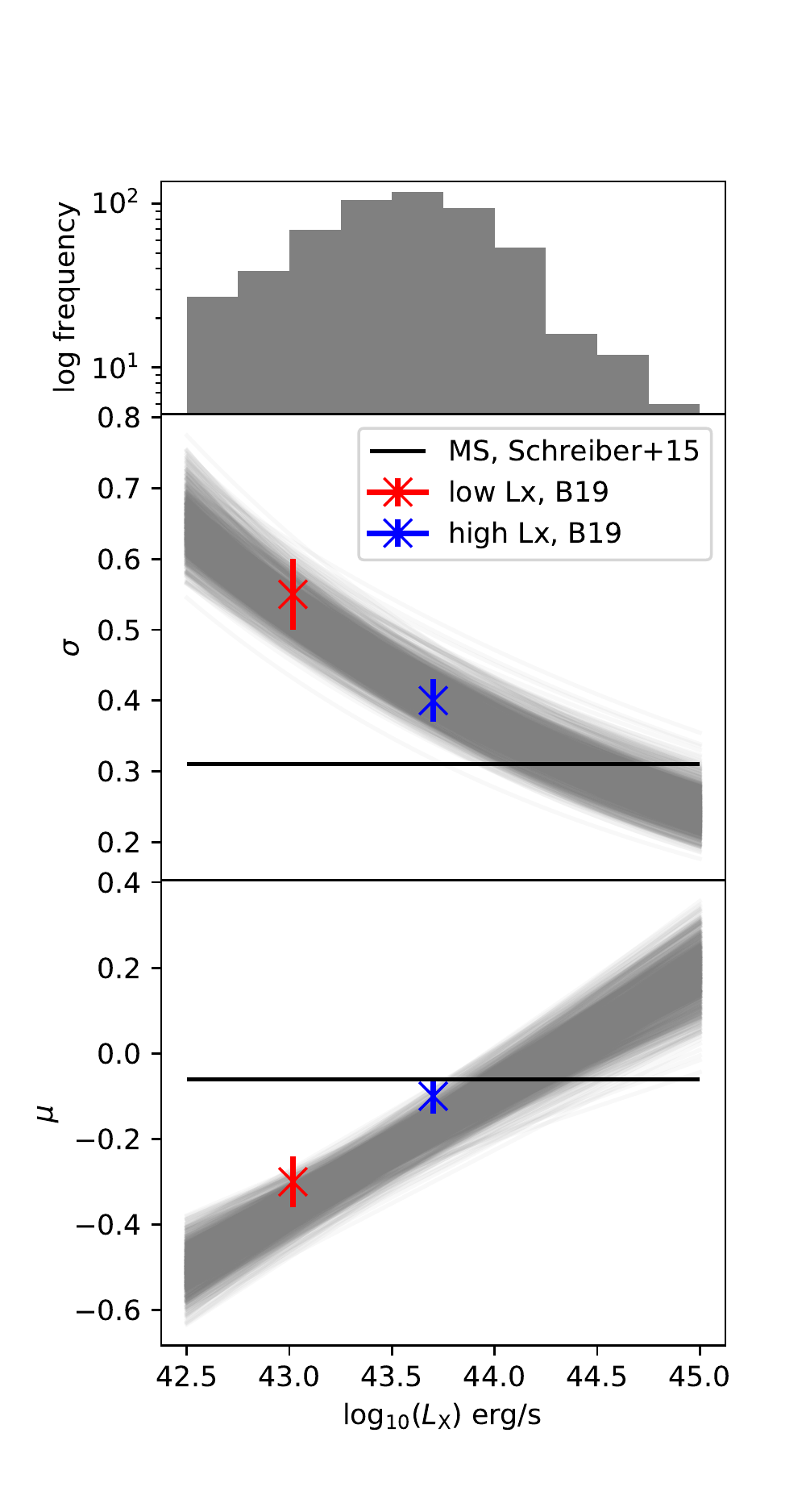}
    \caption{The evolution of the mode, $\mu$, and width, $\sigma$, of the \rms \ distribution as a function of $L_{\rm{X}}$ shown for 1000 bootstrapped samples from the posterior distributions of the hyperparameters, under the assumption of Model 4. Over-plotted are the results from B19, with 1-$\sigma$ errors. Also plotted is the main sequence values from \protect\cite{Schreiber15} (solid black lines). The top plot is the histogram of \lx\ values of the sample for reference.}
    \label{musigmaevol}
\end{figure}

We present summary statistics for the posterior distributions of the hyperparameters in Table~\ref{posttable}. The coefficients of $L_{\rm{X}}$ in the functional relationships (see Equation~\ref{muform}) are given by $\theta_1$ and $\theta_3$, which from Table~\ref{posttable} and Figure~\ref{posterior} are positive and negative respectively. This implies that as $L_{\rm{X}}$ increases, the mode and width of the \rms\ distribution increase and decrease respectively. The relationship between the mode and width of the log-normal \rms\ distribution and $L_{\rm{X}}$ can be seen in Figure~\ref{musigmaevol}, where the posterior distributions of the hyperparameters have been sampled 1000 times and combined with $L_{\rm X}$ to provide samples of $\mu$ and $\sigma$.

\subsection{\rms\ as a function of $L_{\rm X}$}

In this paper, we have used a hierarchical Bayesian framework to remove the need for binning and stacking when modelling the \rms\ distribution of galaxies hosting AGN of different \lx . In doing so, and in contrast to B19, we find \textit{strong} evidence that there is relationship between the \rms\ distribution and \lx\ (i.e., AGN power) as oppose to just tentative evidence.

\begin{figure}
    \includegraphics[width=\linewidth]{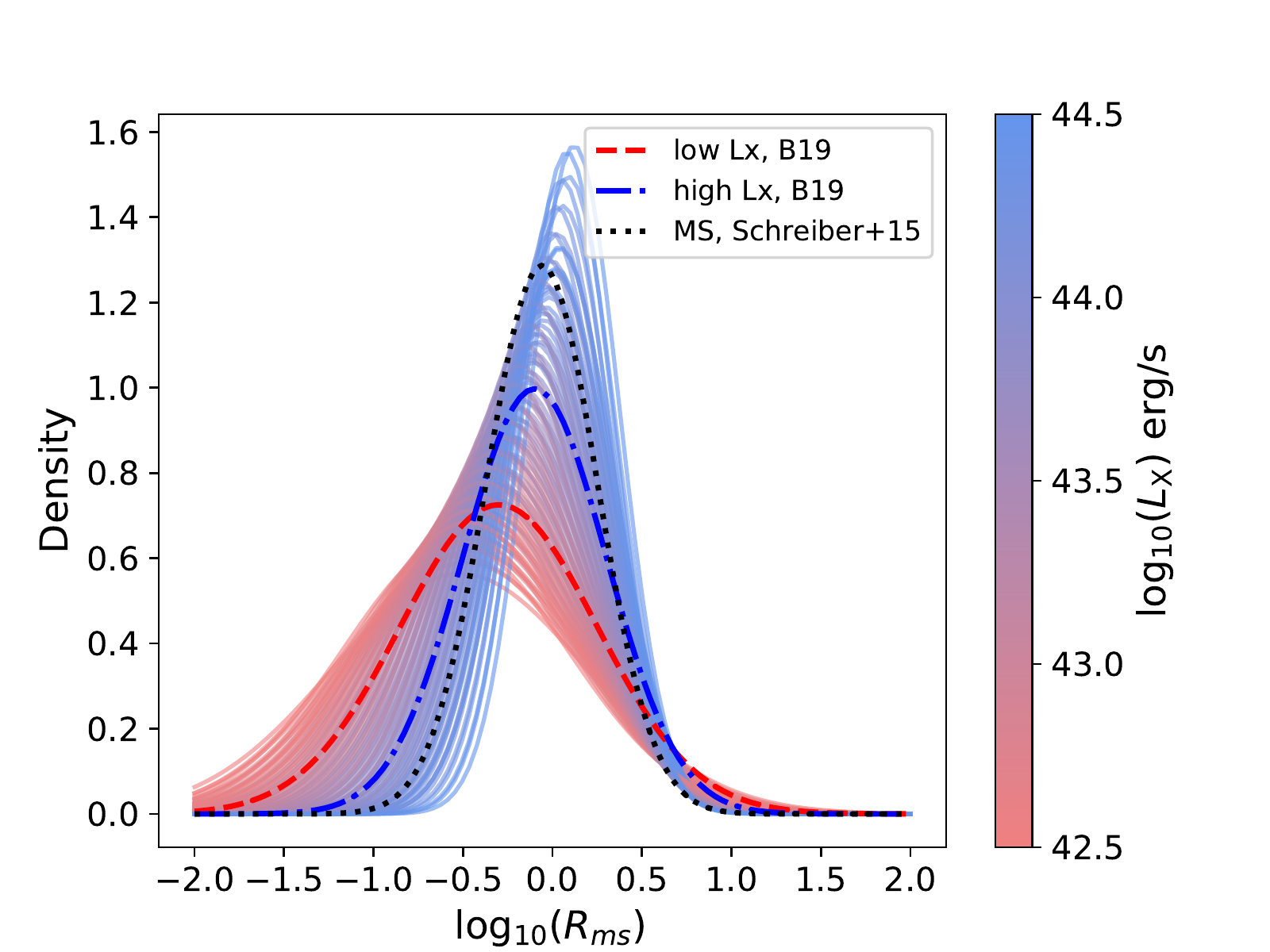}
    \caption{The evolution of the \rms \ distributions as a continuous function of X-ray luminosity, plotted as thin curves. Over plotted are the results from B19 and the $R_{ms}$ distribution for main sequence galaxies from \protect\cite{Schreiber15}. As the X-ray luminosity of a galaxy increases, the probability density function for its \rms \ shifts slightly to higher values and the distribution narrows, consistent with the findings of B19.}
    \label{compare}
\end{figure}

In Figure~\ref{compare} we show how the \rms \ distribution, when modelled as a log-normal distribution, changes as a function of \lx\ in the range $42.53 \leq \log_{10}(L_{\rm X}/\text{ergs s}^{-1})\leq 45.09$. As \lx\ increases, the mode of the \rms \ distribution increases, whilst the width decreases. This is also shown in Figure~\ref{posterior}, as $\theta_1$ takes positive values (i.e., $\mu$ increases with increasing $L_{\rm X}$) and $\theta_3$ takes negative values (i.e., $\sigma$ decreases with increasing \lx ). These results, albeit with more evidence, are still consistent with the tentative findings of B19, which showed that more luminous X-ray AGNs have \rms \ distributions closer to those of main sequence galaxies compared to lower \lx\ AGNs. This is also consistent with the findings of \cite{Schulze19}, who noticed no difference in the SFR distribution of 20 $z\sim2$ quasars and the SFR distribution of main sequence galaxies.

With our new analysis showing stronger evidence of a dependence of \rms\ on \lx , it is natural to ask whether this is consistent with the observed flat relationship between SFR and \lx\ reported by some other studies \cite[e.g.][]{Rosario12, Stanley15}. We are able to explore this issue by generating synthetic SFRs using our \lx -dependent \rms\ model, together with the measured \lx , redshifts, and stellar masses of our sample. To do this we:

\begin{enumerate}
\item{randomly generate a sample from the joint posterior distribution of the hyperparameters, $\theta_{0}^*, \theta_{1}^*, \theta_{2}^*, \theta_{3}^*.$ This involves taking a random point from each of the off-diagonal plots in Figure~\ref{posterior} (and therefore respecting any correlations between parameters);}
\item{for each of the 541 sources in our sample we use their detected $L_{\rm X}$ values, alongside the aforementioned randomly sampled hyperparameters, to calculate the mode and width of the predicted \rms \ distribution. Recall, we reuse the functional relationships we chose earlier so that we have a predicted mode, $\mu_{\rm{pred}}$ and predicted width, $\sigma_{\rm{pred}}$:
  \begin{equation}
    \begin{split}
    &\mu_{\rm{pred}} = \theta_{0}^* + \theta_{1}^* \log_{10}\left(\frac{L_{\rm X}}{10^{40}}\right)\ \  \text{and}\\
    &\sigma_{\rm{pred}} = e^{\theta_{2}^* + \theta_{3}^*\log_{10}\left(\frac{L_{\rm X}}{10^{40}}\right)}.
    \end{split}
  \end{equation}
  }
\item{we then sample an \rms \ value from the log-normal distribution with the parameters $\mu_{\rm{pred}}$ and $\sigma_{\rm{pred}}$;}
\item{we then repeat steps 1-3 10,000 times so that we have, for each source in our sample, a set of 10,000 predicted \rms \ values constrained by our hyperparameter posterior distributions and the assumption of our functional relationships;}
\item{we next multiply each of the sampled \rms \ values by the corresponding main sequence SFR, calculated by using the stellar masses, redshifts and the prescription from \cite{Schreiber15}. This leaves us with a sample of 10,000 predicted SFRs for each source calculated using our functional relationships and posterior distributions.}

\end{enumerate}

\begin{figure*}
  \includegraphics[width=\linewidth]{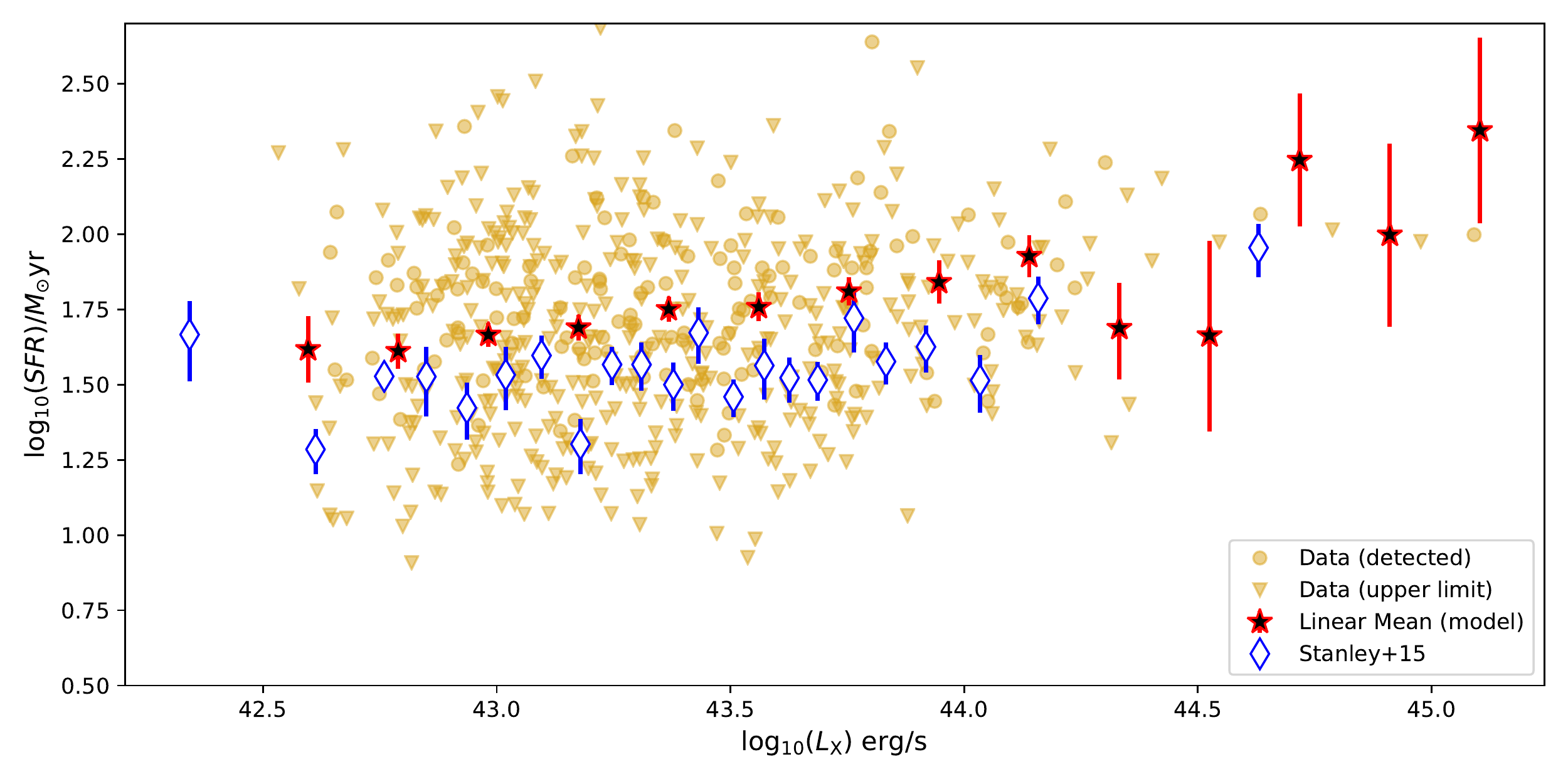}
  \caption{The predicted relationship between SFR and \lx\ using our functional relationships  and hyperparameter posterior distributions. The red stars show the predicted linear mean SFRs in arbitrarily-chosen bins of \lx, calculated using the functional relationships in Equation~\ref{muform}, the main sequence prescription of \protect\cite{Schreiber15} and the stellar mass and redshift of our sources. Also plotted in yellow (circles or triangles) are the SFRs from the raw data (detected and upper limits, respectively). The blue diamonds are the results from \protect\cite{Stanley15} for the redshift range $0.8<z<1.5$, which extends to slightly higher redshifts than our sample. While our results are systematically offset from those of \protect\cite{Stanley15}, they are broadly consistent with their observed flat relationship. We include this plot purely to demonstrate that even after including a significant underlying connection between \rms\ and \lx, the we still obtain a flat relationship between average SFR in bins of \lx.}
  \label{sfrvslx}
\end{figure*}

Figure~\ref{sfrvslx} shows the relationship between SFR and $L_{\rm X}$ as predicted by our \lx -dependent \rms\ distribution. The red stars show the mean predicted SFR in bins of \lx , using a bin width of 0.25 dex (with error bars indicating the $3\sigma$ standard error). Over-plotted are the observed mean SFRs (calculated using survival analysis), also in bins of \lx , from \cite{Stanley15}. The yellow circles represent the SFRs of the 148 AGNs in our sample with measured fluxes, while the yellow triangles represent the upper-limits on SFRs for the remaining 393 AGNs. Despite our analysis providing strong evidence of a relationship between the \rms \ distribution and $L_{\rm X}$, the projected relationship between the average predicted SFRs and $L_{\rm X}$ is comparable to the observed flat relationship of \cite{Stanley15} (i.e., while the means are offset, they are well within the range of scatter given by the observed measurements). While the incorporation of mass and redshift information to convert our predicted \rms\ values to SFR may contribute to some of the flattening, it is plausible that averaging over a log-normal distribution within a particular \lx\ bin could have significantly flattened the relationship also. This further demonstrates that even if a strong underlying relationship between star-forming properties and AGN power exists, it is extremely difficult to extract using average (or even individually-measured) SFRs in bins of \lx.

\begin{table}
    \centering
    \caption{Posterior mean and standard deviations for the hyperparameters for Model 4.}
    \begin{tabular}{cccll}
    \multicolumn{1}{|c|}{Parameter} & \multicolumn{1}{c|}{Mode} & \multicolumn{1}{c|}{Standard Deviation} &  &  \\ \cline{1-3}
    $\theta_0$                      & -1.191                   & 0.119                                   &  &  \\
    $\theta_1$                      & 0.276                     & 0.033                                   &  &  \\
    $\theta_2$                      & 0.540                    & 0.128                                   &  &  \\
    $\theta_3$                      & -0.391                    & 0.040                                   &  &
    \label{posttable}
    \end{tabular}
    \end{table}

\section{Discussion} \label{discussion}

\subsection{Limitations of our approach}

Before discussing the implications of our results, in this Section we aim to highlight limitations of our approach and discuss areas for potential improvement. Initially, as we reuse the same dataset as B19, we have adopted the same set of initial assumptions as that paper. Namely, the assumption about the parametric form of the \rms\ distribution and the validity of the \cite{Schreiber15} main sequence. However by removing the need for binning, we have relaxed the unstated assumption about sources in the same bins having similar properties. The remainder of this Section, therefore aims to highlight additional limitations and assumptions with our methodology, as well as those of B19.

Firstly, the analysis is computationally expensive. This is mostly due to the large number of sampled parameters. In this case, there are four hyperparameters ($\theta_0,...\theta_3$) plus, as described in Section~\ref{priorandpost}, 541 $L_{\rm X}$ parameters with a well-defined (i.e., using by the measured value and its uncertainties) prior distribution. The parameters are sampled pair-wise throughout the MCMC algorithm, which reduces the time, but the algorithm is still computationally expensive. Despite having a large number of parameters, overparameterisation is not a concern since the priors tightly constrain the \lx\ values.

Secondly, in this work, we have imposed simple relationships between the mode and width ($\mu, \sigma$, respectively) of the \rms\ distribution and $L_{\rm X}$. Whilst this relationship could be made more flexible, the aim of this paper was to test the framework and to determine if there is any dependence on $L_{\rm X}$. We therefore chose simple relationships to assess whether we could rule-out the independent case. In future studies, more flexible forms of the functional relationships could be tested and model comparison methods used to determine whether any other functional forms provide a better representation of the data. In addition to making the functional relationships more flexible, other independent variables could be added (such as redshift and stellar mass). By doing so, and allowing for more models to be compared, future studies could use the techniques in this paper to probe deeper into the connection between AGN power and host galaxy properties. As a result of this paper only investigating how the \rms\ distribution changes as a function of \lx, we were cautious that, if there was a significant, systematic change of \lx with redshift, then a redshift evolution in both \lx\ and \rms\ may introduce a spurious positive trend. However, we see no evidence of a strong systematic change of \lx\ with redshift. The median and standard deviation of \lx\ for the lowest and highest redshift quartiles were (43.23, 0.40) and (43.43, 0.44) respectively. Therefore we have no reason to believe that our results are being affected by an underlying redshift evolution in both \lx\ and \rms\ across our redshift bin. With regards to redshift and stellar mass effects, it may be interesting to investigate whether assuming alternative models for the redshift and mass evolution of the Main Sequence \cite[e.g.][]{Speagle14, Ilbert15,Whitaker15,Popesso19a} has a large effect on the results.

Thirdly, posterior model probabilities can be dependent upon the choice of prior distribution chosen for individual parameters. As the marginal likelihood is the integral of the likelihood function over all the prior space (effectively a weighted average of the likelihood function), an analysis of this sort must make sure that the prior distributions are reflective of current up to date knowledge. Our prior distributions are influenced by the work of B19. By the construction of the marginal likelihood, however, overly vague prior distributions can excessively ``penalise'' more complex models. Likewise, prior distributions that are too constrained can favour more complex models. Therefore, prior distributions should be carefully chosen and justified.

Finally, we stress again that we have worked under the assumption that \rms \ distribution is log-normal. This is unlikely to be the case. Indeed, it is known that some AGNs reside in quiescent and starburst galaxies whose combined \rms\ values do not follow a log-normal distribution (e.g. the main sequence/starburst population is believed to follow a bi-modal log-normal distribution in \rms ). Having said that, our focus here is to assess whether, after eliminating the need for binning and averaging (and comparing to the same dataset in B19), the \rms\ distribution could be \lx -dependent. It is not immediately clear why a truly \lx -independent \rms\ distribution would be better modelled by a \lx -dependent log-normal, as opposed to a \lx -independent one. Therefore, we stress we are not suggesting that our model represents the true \rms\ relationship, but instead that an \lx -dependent model is strongly favoured when compared to an \lx -independent one.

\subsection{Implications of our analysis}

The aim of this paper was to introduce a Bayesian hierarchical framework that removes both the need to bin data (particularly in distribution-style analyses) and the need to use averaging techniques (or other summary statistics/parameters). To allow us to accurately demonstrate that any new results were driven by the methodology, we applied our hierarchical model on the same dataset as B19. The process involves assuming a distributional form for one variable (in this case the starburstiness of a galaxy) and setting a direct dependence between the parameters of this distribution and some independent variable (in this case, \lx ). Uncertainties on the independent variable are also fully considered by treating them as a parameter and applying an informative prior, which is derived from the measured values and their uncertainties.

Our results show that, under the assumption that \rms\ is log-normally distributed, there is a strong evidence of a relationship between \rms\ and \lx within the redshift range $0.8<z<1.2$. This reaffirms, to a stronger degree of significance, the result of B19, such that as \lx\ increases, the \rms \ distribution is centered at a higher value and the diversity of \rms \ values decreases. What this implies is that, within the constraints of our model, an \lx $= 10^{44}~{\rm erg~s^{-1}}$ AGN is 21 per cent more likely to reside in a galaxy with \rms$>2$ than an \lx $= 10^{43}~{\rm erg~s^{-1}}$ AGN. This is in agreement with other studies that suggested there is a \textit{tighter} (i.e., more consistent) connection between more luminous AGNs and star formation than for lower-luminosity AGNs \citep[e.g.][]{Rosario13,Stanley17,Aird17,Dai18, Masoura18, Aird19}: for example, it may be that any luminous AGN activity occurs close in time to the star formation activity while lower-luminosity AGN activity can occur when the galaxy is more quiescent (and hence the broader \rms \ distribution) in addition to occurring during the periods of star-formation activity.

In this study, we have investigated the relationship between the \rms\ distribution of AGN hosts and \lx , and found strong evidence of a relationship between the two. Recently, a number of studies have approached this problem from the other direction; i.e., investigating how AGN power changes as a function of the star-forming properties of their hosts. For example, \cite{Chen13} reported that, when binned in terms of SFR, the mean \lx\ of star-forming galaxies increases with average SFR (see also \citealt{Delvecchio15}, who also accounted for the effects of galaxy stellar mass). Further, \cite{Rodighiero15} found that, when binning according to stellar mass, the mean \lx\ of starburst galaxies is higher then that of main sequence galaxies which, in turn, is higher than that of quiescent galaxies. Both these results imply that average AGN power is higher in more actively star-forming systems. More recently, \cite{Aird19} and \cite{Grimmett19} have shown that the distribution of specific \lx\ (i.e., $=L_{\rm X}/M_*$, a proxy for Eddington ratio $\lambda_{\rm Edd}$), changes as a function of the star-forming activity of their hosts, with a higher fraction of starbursts hosting AGNs with $\lambda_{\rm Edd}>10\%$ than their main sequence counterparts. By exploring how the star-forming properties of galaxies change as a function of \lx , this study (and B19) take the opposite approach. While there are significant differences between the properties being considered in each study (not least the exploration of Eddington ratio in \citealt{Aird19} and \cite{Grimmett19}, whereas we only consider \lx\ here) all appear to support the assertion that more powerful AGNs (whether expressed in terms of \lx\ or Eddington ratio) are preferentially found in more actively star-forming systems.

\section{Conclusions}

In this work we have introduced a hierarchical Bayesian framework to assess whether the \rms\ distribution of AGN-hosting galaxies changes as a \textit{continuous} function of an X-ray luminosity (\lx ). Our approach removes the need for both binning and averaging and also allows for full consideration of the uncertainties on the independent variable.

By modelling the \rms\ distribution as a log-normal, and proposing simple relationships between its parameters (i.e., mode and width) of that log-normal and X-ray luminosity, we found strong evidence that an \lx -dependent model is preferred over an \lx -independent one. By binning the same data, B19 reported the same overall trend, but without such strong evidence, thereby highlighting the importance of utilising all available information by removing the need for binning. By using the same dataset and pre-processing as B19, we ensured that any differences found in contrast to that paper are a direct result of the new analysis technique.

Despite finding a strong relationship between the \rms\ distribution and AGN power, when we convert our \lx-dependent distributions back into the mean SFR - \lx\ plane, we find that the \textit{dependent} model can reproduce results consistent with previously seen flat relationships \cite[e.g.][]{Stanley15}. This further highlights the difficulty in extracting underlying relationships between AGN power and host galaxy properties when averaging in bins of AGN power.
 
\section*{Acknowledgements}

We would like to thank the anonymous referee for their prompt and detailed report which improved the quality of the paper. LG is supported through a PhD scholarship granted by the University of Sheffield. EB and JM acknowledge STFC grant R/151397-11-1. Throughout the preparation of this manuscript, we made extensive use of the Numpy, Scipy \citep{Virtanen20} and Matplotlib \citep{Hunter07} packages.  

\pagebreak



\bibliographystyle{mnras}
\bibliography{/local/php16lpg/bibtex/ref} 




\appendix
\section{The full MCMC sampler}\label{app:mcmc}
\begin{table*}
    \begin{tabular}{|c|c|c|c|c|c|c|}
        \hline
        Case               & Current $\theta_1$                   & Proposed $\theta_1'$                  & Model now & Model proposed & $q(\theta, \theta')$                                                & $q(\theta', \theta)$                                                \\ \hline
        \multirow{2}{*}{A} & \multirow{2}{*}{$\theta_1 = 0$}      & \multirow{2}{*}{$\theta_1' = 0$}      & 1         & 1              & \multirow{2}{*}{$0.5\times f(\theta_0'|\theta_0, s_1)$}                                                & \multirow{2}{*}{$0.5\times f(\theta_0|\theta_0', s_1)$}                                                \\
                           &                                      &                                       & 3         & 3              &                                                                     &                                                                     \\ \hline
        \multirow{2}{*}{B} & \multirow{2}{*}{$\theta_1 = 0$}      & \multirow{2}{*}{$\theta_1' \neq 0$}   & 1         & 2              & \multirow{2}{*}{$0.5\times f_2(\: [\theta_0', \theta_1']\:|\: [\hat{\theta_0}, \hat{\theta_1}], \Sigma_1)$}          & \multirow{2}{*}{$0.5 \times f(\theta_0|\hat{\theta_0}, s_2)$}                                                \\
                           &                                      &                                       & 3         & 4              &                                                                     &                                                                     \\ \hline
        \multirow{2}{*}{C} & \multirow{2}{*}{$\theta_1 \neq 0$} & \multirow{2}{*}{$\theta_1' = 0$}      & 2         & 1              & \multirow{2}{*}{$0.5 \times f(\theta_0|\hat{\theta_0}, s_2)$}                                                & \multirow{2}{*}{$0.5\times f_2( \:[\theta_0, \theta_1]\:| \:[\hat{\theta_0}, \hat{\theta_1}], \Sigma_1)$}           \\
                           &                                      &                                       & 4         & 3              &                                                                     &                                                                     \\ \hline
        \multirow{2}{*}{D} & \multirow{2}{*}{$\theta_1 \neq 0$} & \multirow{2}{*}{$\theta_1' \neq 0$} & 2         & 2              & \multirow{2}{*}{$0.5\times f_2(\: [\theta_0', \theta_1']\:|\: [\theta_0, \theta_1], \Sigma_2)$} & \multirow{2}{*}{$0.5\times f_2(\: [\theta_0, \theta_1]\:|\: [\theta_0', \theta_1'], \Sigma_2)$} \\
                           &                                      &                                       & 4         & 4              &                                                                     &                                                                     \\ \hline
        \end{tabular}
        \caption{Summary of the possible model switches for 1 proposal of the $\mu$-related hyperparameters, $\theta_0$ and $\theta_1$. There are four potential cases depending on whether the model is currently in a $\mu$-dependent or a $\mu$-independent state and whether we propose to move to a $\mu$-dependent or $\mu$-independent state. For the possible cases the value of the proposal density $q(\theta, \theta')$ and the inverse $q(\theta', \theta)$ are given. The univariate Gaussian density is given by $f$ and the bivariate Gaussian density is given by $f_2$. The tuned proposal widths are given by $s_1$ and $s_2$, and the calculated covariance matrices by $\Sigma_1$ and $\Sigma_2$. When a model switch is proposed, the ``reasonable'' values must be used to sample a proposed parameter value and these are given by $\hat{\theta_0}$ and $\hat{\theta_1}$.}
        \label{app:tab}
\end{table*}

In this appendix we describe, in detail, one full step of the MCMC sampler used to construct the posterior distributions presented in Section~\ref{results}, which were then used to compare our various models. Interested readers should also refer to the study of \cite{Gottardo08}, from which our sampler is adapted.

A key component of our algorithm is that, when it proposes a switch between models, it proposes ``reasonable'' parameters within the proposed model. Otherwise, we run the risk of never switching models -- not because the proposed model is necessarily worse, but because we always propose highly unlikely parameter values within that model. What we mean by ``reasonable'', therefore, is likely parameter values within each proposed model. As such, we need to have some knowledge of the posterior probability distributions of each model before we can start proposing switches between models. One way of achieving this would be to force Model 1, for example, to converge, then force a switch to Model 2, allow that to converge, and so on. Once all models have converged, we would then allow our sampler to switch between models by proposing reasonable parameter values (i.e., those close to the posterior mode). In our case, however, as we only have four models, we instead run a separate standard MCMC sampler for each model (i.e., without model switching), which gives us an indication of the most suitable regions of the posterior parameter space for each model. Mathematically, these two approaches are exactly analogous.

With an estimate of the posterior parameter space for each model in-hand, we can propose reasonable regions of the parameter space when switching between models. In what follows, we describe how we switch between various models. For ease of explanation, we will only consider $\theta_0$ and $\theta_1$, but same process is applied when sampling $\theta_2$ and $\theta_3$. Recalling that we step through the parameters in pairs, we sample $\theta_0$ and $\theta_1$ at the same time. This leads to four possible cases, which are summarised in Table \ref{app:tab}, and discussed in detail below.

{\bf Case A:} Here, the sampler is currently in the state where $\theta_1=0$, and is proposing $\theta_1=0$ (i.e., it is in a $\mu$-independent model $[$Models 1 or 3$]$ and proposes to remain within a $\mu$-independent model). However, because we progress through the vector pairwise, the sampler must still propose a $\theta_0$ value. For this, we use a standard MH proposal -- a value randomly selected from a Gaussian distribution centered on the current $\theta_0$ value. Based on pilot runs, we choose a value for the width of the Gaussian distribution that results in good mixing (i.e., the acceptance rate is between 20--40 per cent). In this case, the $q(\theta,\theta')$ value is the product of the likelihood of choosing $\theta_1'=0$ (i.e., 0.5) and the proposed $\theta_0$ value (i.e., $\theta_0'=f(\theta_0'|\theta_0, s_1)$, where $f$ is the Gaussian density function). This product is symmetrical on switching between $\theta$ and $\theta'$, meaning $q(\theta,\theta')=q(\theta',\theta)$, so the $q$ terms cancel in Equation~\ref{ar}.

{\bf Case B:} In this case, the sampler is currently in the state where $\theta_1=0$, and is proposing $\theta_1\neq0$ (i.e., it is in a $\mu$-independent model $[$Models 1 or 3$]$ and is proposing to switch to a $\mu$-dependent model [Models 2 or 4]). As a result of proposing a switch to a $\mu$-dependent model, we must propose values for both $\theta_0$ and $\theta_1$. To do this, we use a bivariate Gaussian distribution, centered on the ``reasonable'' values obtained using the process described above. Based on pilot runs, we choose a value for the widths of the bivariate Gaussian distribution that results in good mixing (i.e., the acceptance rate is between 20--40 per cent). In addition to the widths, the bivariate Gaussian distribution accounts for the correlation between $\theta_0$ and $\theta_1$ by using the calculated covariance matrix. In this case, the $q(\theta,\theta')$ value is the product of the likelihood of choosing $\theta_1'\neq0$ (i.e., 0.5) and the proposed $\theta$ values (i.e., $\theta'=f_2(\:[\theta_0', \theta_1']\:|\:[\hat{\theta_0},\hat{\theta_1}], \Sigma_1)$, where $f_2$ is the bivariate Gaussian density function, $\hat{\theta_0}, \hat{\theta_1}$ are the estimates of the posterior mode from the original chains and $\Sigma_1$ is the covariance matrix. This product is not symmetrical on switching between $\theta$ and $\theta'$, since the inverse process involves sampling from a univariate Gaussian. This means $q(\theta,\theta') \neq q(\theta',\theta)$, so they must be accounted for in the acceptance ratio.

{\bf Case C:} Here, the sampler is currently in the state where $\theta_1\neq0$, and is proposing $\theta_1'=0$ (i.e., it is in a $\mu$-dependent model $[$Models 2 or 4$]$ and is proposing to switch to a $\mu$-independent model [Models 1 or 3]). As a result of proposing a switch to a $\mu$-independent model, we again must propose a ``reasonable'' value of $\theta_0$ within the proposed model. To do this, we use a distribution, centered on the ``reasonable'' values obtained using the process described above. Based on pilot runs, we choose a value for the width of the Gaussian distribution that results in good mixing (i.e., the acceptance rate is between 20--40 per cent). In this case, the $q(\theta,\theta')$ value is the product of the likelihood of choosing $\theta_1'=0$ (i.e., 0.5) and the proposed $\theta_0$ value (i.e., $\theta_0'=f(\theta_0'|\hat{\theta_0}, s_2)$, where $f$ is the Gaussian density function, $\hat{\theta_0}, \hat{\theta_1}$ are the estimates of the posterior mode from the original chains and $\Sigma_1$ is the covariance matrix). This product is not symmetrical on switching between $\theta$ and $\theta'$ for the same reason as in Case B (i.e., the inverse process involves sampling from a bivariate Gaussian distribution). This means $q(\theta,\theta') \neq q(\theta',\theta)$, so they must be accounted for in the acceptance ratio.

{\bf Case D:} In this final case, the sampler is currently in the state where $\theta_1\neq0$, and is proposing $\theta_1'\neq0$ (i.e., it is in a $\mu$-dependent model $[$Models 2 or 4$]$ and is proposing to remain in a $\mu$-dependent model). As a result we need to propose values for both $\theta_0$ and $\theta_1$. To do this, we use a bivariate Gaussian distribution, centered on the current values. Based on pilot runs, we choose a value for the width of the Gaussian distribution that results in good mixing (i.e., the acceptance rate is between 20--40 per cent) and calculate the appropriate covariance matrix. In this case, the $q(\theta,\theta')$ value is the product of the likelihood of choosing $\theta_1'\neq0$ (i.e., 0.5) and the proposed $\theta$ value (i.e., $\theta'=f_2(\:[\theta_0', \theta_1']\:|\:[\theta_0,\theta_1], \Sigma_1)$, where $f_2$ is the bivariate Gaussian density function, and $\Sigma_2$ is the covariance matrix). This product is symmetrical on switching between $\theta$ and $\theta'$, meaning $q(\theta,\theta') = q(\theta',\theta)$ and so the terms cancel.

This process is then repeated for the next pair of hyperparameters (i.e., $\theta_2$ and $\theta_3$) followed by one sampling through the \lx\ values individually (i.e., not pair-wise), the latter of which is done by using a standard MH algorithm.


\bsp    
\label{lastpage}
\end{document}